\documentstyle[12pt]{article}

\makeatletter

\@addtoreset{equation}{section}
\def\section{\@startsection {section}{1}{\z@}{-3.5ex plus -1ex minus
 -.2ex}{2.3ex plus .2ex}{\large\bf\centering}}
\def\subsection{\@startsection{subsection}{2}{\z@}{-3.25ex plus
 -1ex minus -.2ex}{1.5ex plus .2ex}{\sc}}
\fnsymbol{footnote}

\advance \voffset by -1.0in
\advance \hoffset by -0.6in
\def\cd{\!\cdot\!}

\def\bea{\begin{eqnarray}}
\def\eea{\end{eqnarray}}
\def\bphi{{\mbox{\boldmath $\phi$}}}
\def\bn{{\mbox{\boldmath $n$}}}
\def\bl{{\mbox{\boldmath $l$}}}
\def\bE{{\mbox{\boldmath $E$}}}
\def\bx{{\mbox{\boldmath $x$}}}

\begin{document}
\baselineskip 18pt
\parskip 5pt
\begin{flushright}

DTP 95-27

hep-th/9506004

30 May 1995

\end{flushright}

\vspace {2cm}

\begin{center}

{\LARGE Bogomol'nyi solitons in a gauged $O(3)$ sigma model}

\vspace{1cm}
\baselineskip 24pt

{\Large
B.J. Schroers\footnote{e-mail: {\tt b.j.schroers@durham.ac.uk}}
\footnote{address from 1 September 1995: Faculty of Physics and
Astronomy, University of Amsterdam,
Valckenierstraat 65, 1018 XE Amsterdam,
The Netherlands}
\\
Department of Mathematical Sciences, South Road
\\
Durham DH1 3LE, United Kingdom}
\\
\baselineskip 18pt

\vspace{1.5cm}

{\bf Abstract}

\end{center}
\baselineskip 15pt

{\small
\noindent The scale invariance of the $O(3)$ sigma model can be
broken by gauging  a $U(1)$ subgroup of the $O(3)$ symmetry
and including a Maxwell term for the gauge field  in the
Lagrangian. Adding also a suitable potential  one obtains
a  field theory of Bogomol'nyi type with topological solitons.
These solitons  are stable against  rescaling  and carry magnetic
flux which  can take arbitrary values in some finite interval.
The soliton mass is independent of the flux, but the soliton
size depends on it. However, dynamically changing the flux requires
infinite energy, so the flux, and hence the soliton size,
remains constant during time evolution. }

\vspace{2cm}
\noindent {\it keywords}: sigma  models;  Bogomol'nyi equations;
 Skyrme-Maxwell theory

\noindent PACS number(s): 11.10.Kk, 11.10.Lm,  11.27.+d

\vfill

\pagebreak

\baselineskip 18pt

\section{Introduction}

The $O(3)$ sigma model in (2+1) dimensions is a popular
model in theoretical physics. Statically it is integrable
and of  Bogomol'nyi type, i.e. all minimal energy
solutions can be obtained by solving the  first order Bogomol'nyi
equations (which imply the second order Euler-Lagrange  equations).
As a result  one can explicitly  write down soliton solutions of
arbitrary degree in term of rational functions \cite{BP}.
 From the point of view of a particle physicist,  however,
the model has one  important drawback: it is scale invariant
and as a result its soliton solutions  have an arbitrary size,
making them unsuitable as  models for particles.
Numerical simulations of the  solitons'  interaction behaviour
in the (2+1)-dimensional model suggest that the solitons do
indeed change their size during interactions and generically
turn into  singular configurations of zero size \cite{spike}.
The most obvious way to break the scale invariance of the
model is  to add terms to the Lagrangian density which
contain a different number of derivatives from the sigma model
term (which contains two). Indeed, the inclusion of a Skyrme term
(four derivatives) and a potential term (no derivatives) leads
to so-called baby Skyrme models which have soliton solutions
of definite size. Such models
are close analogues of the (3+1) dimensional Skyrme model
and therefore physically interesting, but they
 are neither integrable nor of
Bogomol'nyi type and  can only be studied with considerable numerical
effort \cite{babyskyrme}.
A mathematically more elegant way of breaking the
scale invariance of the $O(3)$ sigma model is to add a potential
term only and prevent the solitons from collapsing by  making
them spin. In \cite{Leese1}
 it was shown that with a suitable choice of the potential such
a model is of  Bogomol'nyi type. Its soliton solutions,
called $Q$-lumps, can be written down explicitly and their
interaction behaviour was studied in  \cite{Leese2}.

Here we investigate the possibility of breaking
the scale invariance of the sigma model by
introducing a $U(1)$ gauge field whose dynamics is governed by
a Maxwell term. This possibility is also potentially of  interest
in the  (3+1)-dimensional Skyrme  model.  There the gauging of a  $U(1)$
subgroup and the inclusion of a Maxwell term is physically natural, and
it would be aesthetically appealing if one could do away with the
Skyrme term  and still retain a  model with stable soliton solutions.
Since in three spatial dimensions  the sigma model term  has the
opposite scaling behaviour from  the Maxwell term,
 simple scaling arguments do not rule out
solitons of a definite size in a ``Skyrme-Maxwell model without a
Skyrme  term". At the end of the paper we will make some conjectures
about this possibility.

Since the solitons here carry magnetic flux  it is appropriate
to compare them to vortices. Vortices either have quantised flux
in which case they are topologically stable (e.g. in the abelian
Higgs model, see \cite{bogvor} and references therein)
or they have arbitrary flux in which case they are
not topologically stable (e.g. non-topological Chern-Simons vortices, see
\cite{csvor}).
The topological stability of the  solitons studied here, however,  is
independent of their magnetic  flux. Thus they can  have arbitrary flux
and yet be topologically stable.

\section{Topological stability}

In the $O(3)$ sigma model the basic field is  a map from (2+1)-dimensional
Minkowski space
to the  2-sphere of unit radius. Here  Minkwoski space is assumed to
have  the  signature  $(-,+,+)$,   and its elements are written as
$(t,\bx)$  or alternatively  $x^{\alpha}$, $\alpha=0,1,2$;  for
partial derivatives with respect to these coordinates we write
$\partial_{\alpha}$. The field, denoted $\bphi$,  has three components
$\phi_1,\phi_2$ and $\phi_3$ satisfying the constraint $\bphi\cd \bphi =
\phi_1^2+ \phi_2^2+\phi_3^2 =1$.
The potential energy functional is
\bea
E_{\sigma} [\bphi] =   \frac{1}{2} \int \, d^2 x \,
 \left( (\partial_1 \bphi)^2 + (\partial_2\bphi)^2 \right).
\eea
To obtain finite-energy configurations one requires that,  at all times $t$,
\bea
\label{pbound}
\lim_{|\bx|\rightarrow \infty} \bphi(t,\bx) = \bn,
\eea
where $\bn$ is a constant unit vector which we take to
be $\bn =(0,0,1)$ for definiteness. This condition allows
one to add the point $\infty$ to physical space  {\bf R}$^2$, thus
 compactifying it  to a topological 2-sphere. As a result
a field $\bphi$ at a fixed time may be viewed as a map from one 2-sphere
to another and therefore has an associated degree
deg$[\bphi]$. This degree is a homotopy invariant and therefore cannot
change  during time evolution.

 For our purposes it is convenient to  express
 the degree in terms of the current
\bea
k_{\alpha} =\frac{1}{2} \epsilon_{\alpha\beta\gamma}
 \bphi\cd\partial^{\beta}\bphi
\times \partial^{\gamma}\bphi,
\eea
where
$\epsilon_{\alpha \beta \gamma}$  is  the totally antisymmetric tensor
in three indices, normalised so that $\epsilon_{012}=1$.
One finds
\bea
\mbox{deg}[\bphi]=\frac{1}{8\pi}\int d^2 x \,k_0 = \frac{1}{4\pi}
 \int d^2 x \,
\bphi\cd \partial_1 \bphi
\times\partial_2\bphi.
\label{Q}
\eea
 It is
easy to see that the divergence of  $k_{\alpha}$
vanishes independently of the equations of motion; together
 with the boundary
condition (\ref{pbound})
this explicitly shows the conservation of the degree.
The  degree of a configuration  is also important because it provides
a lower bound on   its energy \cite{BP}:
\bea
\label{bog}
E_{\sigma}[\bphi] \geq 4\pi \, |\mbox{deg}[\bphi]| \,.
\eea

The energy functional $E_{\sigma}$ and
the boundary condition (\ref{pbound}) are  invariant under the
 group of global rotations of the field $\bphi$ about
 the  fixed vector $\bn$. This is
  the $U(1)$ symmetry we want to
gauge. Thus we introduce a  $U(1)$ gauge field $A_{\alpha}$ and a covariant
derivative
\bea
D_{\alpha} \bphi  = \partial_{\alpha}\bphi + A_{\alpha} \bn \times \bphi.
\eea
Defining the   field strength as usual via $F_{\alpha \beta}
=\partial_{\alpha} A_{\beta} - \partial_{\beta} A_{\alpha}$ with magnetic
field $F_{12}$, we can write
down the potential
energy functional which is the subject of this paper:
\bea
\label{gauge}
E_{\mbox{\tiny gauge}}[\bphi,A_1,A_2] = \frac {1} {2}\int
d^2 x\,   \left (( D_1 \bphi)^2 + (D_2\bphi)^2 +
( 1-{\bf n} \cd \bphi)^2 +
 F_{12}^2\right).
\eea

In the gauged model the topological current $k_{\alpha}$, while still
divergence free,  is unsatisfactory
because it is not gauge invariant. Through trial
and error  one finds that the current
\bea
j_{\alpha} = \frac {1}{2} \epsilon_{\alpha \beta \gamma}
\left( \bphi\cd D^{\beta}\bphi\times D^{\gamma}\bphi + F^{\beta\gamma}
(1-{\bf n}\cd \bphi) \right),
\eea
which is manifestly gauge invariant, differs from $k_{\alpha}$
only by the curl of another vector field
\bea
j_{\alpha} = k_{\alpha} + \epsilon_{\alpha \beta \gamma} \partial^{\beta}
\left((1-{\bf n}\cd \bphi) A^{\gamma} \right).
\eea
Thus $j_{\alpha}$  also has vanishing divergence,
 $\partial_{\alpha} j^{\alpha}=0$, and together with the boundary condition
(\ref{pbound})  this implies that   the  degree of a configuration
$\bphi$ can be expressed as
\bea
\label{gaugeQ}
\mbox{deg}[\bphi] = \frac{1} {8\pi} \int d^2 x \,j_0
=\frac{1}{4\pi} \int d^2 x \,\left( \bphi\cd D_1\bphi\times D_2\bphi
 + F_{12} (1-{\bf n}\cd \bphi) \right).
\eea

Returning to the   energy functional $E_{\mbox{\tiny gauge}}$
we note that it  can be re-written
as
\bea
E_{\mbox{\tiny gauge}}[\bphi, A_1,A_2] =  \frac{1}{2} \int
d^2 x\, \left( ( D_1 \bphi \pm \bphi \times  D_2 \bphi )^2
+ ( F_{12} \mp (1- {\bf n}\cd \bphi))^2 \right) \nonumber \\
\pm
\int d^2 x \,\left(  \bphi\cd D_1\bphi   \times D_2 \bphi + F_{12}
(1-{\bf n}\cd \bphi ) \right),
\eea
where we used that $\bphi \cd D_\alpha \bphi=0$.
Together with the formula for  the degree (\ref{gaugeQ}) this implies
\bea
E_{\mbox{\tiny gauge}}[\bphi, A_1, A_2] \geq 4\pi |\mbox{deg}[\bphi]|,
\eea
with equality if and only if one of the
Bogomol'nyi equations holds:
\bea
\label{Bogeq}
D_1\bphi &=& \mp \bphi \times D_2\bphi \nonumber \\
F_{12} &=& \pm (1-{\bf n}\cd \bphi).
\eea
It is instructive to write these equations also
in a different  form which results when the target space
$S^2$ is stereographically
projected onto {\bf C} $\cup \{\infty\}$.  More precisely
  defining  a complex-valued  field $u =u_1 +iu_2$
via
\bea
u_1 = { \phi_1 \over 1+ \phi_3}  \qquad u_2 = {\phi_2 \over 1+ \phi_3},
\eea
the  Eqs. (\ref{Bogeq})  become, in terms of
$u$,
\bea
\label{Bogeq2}
D_1 u &=&  \mp i D_2 u\nonumber \\
F_{12} &=& \pm {2 |u|^2 \over 1+ |u|^2},
\eea
where $D_j$ now stands for $\partial_j + i A_j$, $j=1,2$. In the gauge
$\partial_1 A_1 + \partial_2 A_2 =0$ these equations imply
the following second
order equation for $\varphi = \ln u$:
\bea
\Delta \varphi = {2\over 1 + e^{-(\varphi + \bar \varphi})}.
\eea
The integrability of such ``non-linear Laplace equations" has
been studied in the literature, but the present equation
lies outside a small class of such equations which are
known to be integrable by standard methods, such as
scattering  transforms \cite{nonint}.
To find solutions of the Bogomol'nyi equations (\ref{Bogeq})
 we therefore resort to numerical methods.

\section{Solving the  Bogomol'nyi equations}

When  seeking  solutions of Eq. (\ref{Bogeq}) with non-zero degree
we  restrict attention to fields which are invariant under
simultaneous rotations and reflections in space and target  space.
Thus we assume that $\bphi$ is of the  so-called hedgehog form
\bea
\label{hedge}
\bphi(\bx) = (\sin f(r) \cos N\theta, \sin f(r) \sin N\theta
, \cos f (r)),
\eea
where $(r,\theta)$ are polar coordinates in the $\bx$-plane,
$N$ is a non-zero integer and $f$ is a  function satisfying
 certain boundary conditions to be specified below.
The gauge field is assumed to have only a $\theta$-component which is
 of the form
\bea
A_{\theta} =  N a(r).
\eea
(For a more detailed justification of this ansatz see \cite{MaxSkyr}.)
Then the magnetic component of the field strength is simply
\bea
F_{12} = N{a' \over r}.
\eea
To obtain  fields which  are  regular at the origin we require
\bea
\label{origin}
f(0) = \pi \quad \mbox{and} \quad a(0) =0,
\eea
and to ensure also that the energy  is finite
we   impose
\bea
\label{infty}
\lim_{r\rightarrow \infty} f(r) = 0
 \quad \mbox{and} \quad \lim_{r\rightarrow
\infty} a'(r) = 0\,.
\eea
One checks that the degree of such a configuration is $-N$.

The Bogomol'nyi equations (\ref{Bogeq}) imply the following coupled first
order
differential equation for $f$ and $a$:
\bea
f' &=& - |N| {a+1 \over r} \sin f \label{f} \\
a' &=& - {r\over |N|} (1-\cos f)\, , \label{a}
\eea
where the  alternative signs of (\ref{Bogeq}) have been absorbed into
the modulus sign.
For brevity we will refer to  the boundary value problem posed by
these differential equations together with the  boundary conditions
(\ref{origin}) and (\ref{infty}) as BVP. As a first step in its
discussion we establish the

\noindent {\bf Proposition}:
{\it  The boundary value problem BVP  has  no solution if} $|N| =1$,
 {\it but  it has  a one-parameter family
of solutions  if} $|N|>1$.

Before  entering the proof   we
 note the solutions of  Eqs. (\ref{f}) and
(\ref{a}) for small $r$.  Using the  boundary condition (\ref{origin})
 and keeping only the leading powers in $r$ one  finds
\bea
\label{small}
f \approx \pi + Ar ^{|N|} \quad \mbox{and }
 \quad a \approx -{1\over 2|N|} r^2,
\eea
where $A$ is an {\it arbitrary} constant. When   integrating  Eqs.
(\ref{f}) and (\ref{a})  numerically one cannot start the
integration at  the regular-singular point $r=0$. Instead we integrate
from some small value ($r=10^{-6}$  in practice) outwards, imposing
the initial values there according to (\ref{small}).  According to
the proposition, there is a family of  values for
 $A$ that will lead to a solution
satisfying the boundary conditions (\ref{infty}) at infinity when $|N|>1$
but  there is  no such value when $|N|=1$.

The proof of the proposition proceeds in  four steps; in it a function
$g$ is called increasing (decreasing) if
 $x>y \Rightarrow g(x) \geq  g(y)$ ($x>y \Rightarrow g(x) \leq  g(y)$).

\noindent {\bf 1}.
It is clear from  Eq. (\ref{a})
that  $a' \leq 0$, i.e.  $a$ is a decreasing function. Moreover,
since $0\leq (1 -\cos f)
\leq 2$, the boundary conditions (\ref{origin}) and the intermediate
value theorem imply
\bea
\label{aineq}
-{r^2\over 2 |N|} \leq a < 0.
\eea

\noindent {\bf 2}.
For any solution of BVP,  $f(r) \in (0,\pi]$ for all $r$.
To  prove this
  we  first show that  $f(r) > 0$ for all $r$.
For suppose that $f$ were close to zero,
so that  Eqs. (\ref{f}) and (\ref{a})
simplify to
\bea
f' &=& - |N| {a+1 \over r}  f \label{smallf} \\
a' &=& - {r\over 2 |N|}  f^2 . \label{smalla}
\eea
Then, from the first equation $\int  d\, \ln f = -|N| \int dr \, (a+1)/r$.
However, while the  left hand side diverges as $f\rightarrow 0$, the
right hand  side is finite for any finite interval of integration
(we can assume without loss of generality that $r=0$ is not included in
 the interval).  Thus $f$ cannot vanish for any finite value of $r$;
since it is continuous and non-zero for $r=0$ it is positive for all $r$.
By a similar argument one shows that if $f(r_0) > \pi$ for some
$r_0 >0$ then  $f(r)\ > \pi $ for all $r>r_0$, which  violates
the boundary condition (\ref{infty}). Thus  $0 < f\leq \pi$ as claimed.

\noindent {\bf 3}. For any solution of BVP, $a(r)  > -1$ for all $r$.
Suppose this were not the case. Then there is an $r_1$  such that
$a(r_1) \leq  -1$  and, since $a$ is decreasing,
$a(r) \leq -1 $ for all $r>r_1$.
 However, since $f\in (0,\pi]$ , Eq. (\ref{f}) then implies  that $f$ is
increasing for $r>r_1$, which is incompatible with the boundary
condition (\ref{infty}). Combining this result  with the inequality
 (\ref{aineq}) we conclude  $-1 <  a < 0$. Note that
 it  then follows   from Eq. (\ref{f}) that $f$ is a
decreasing function.

\noindent {\bf 4}.
 Since  $a$ is  bounded below by $-1$ and  decreasing
the limit $\lim_{r \rightarrow
\infty} a(r) = \alpha$ exists and $\alpha \in [-1,0)$.
For large $r$, and hence small $f$,  Eq. (\ref{smallf})  holds and
 implies that $f$ is asymptotic
 to $Cr^{-|N|(\alpha +1)}$, where $C$ is some
positive constant. However, Eq. (\ref{smalla}) then tells us that for
large $r$, $a' \approx -(C^2/2|N|)r^{(1-2|N|(\alpha+1))}$.
Thus $a$ can only converge if $2|N|(\alpha+1) - 1 >1$,
or equivalently
\bea
\alpha > \frac{1} {|N|} -1.
\eea
This is impossible to satisfy if $N=1$, but for $|N|>1$ there is whole
interval $(-1+1/|N|, 0)$ of acceptable asymptotic values for $a$, and
hence there is  a  one-parameter family of solutions of BVP,
 which was to be shown.

When $|N|>1$ the variable parametrising the solutions of BVP
can be taken to be  $\alpha$ or,
more physically, the total magnetic flux $\Phi$
which is related  to $\alpha$
via
\bea
\Phi = \int d^2 x \, F_{12}= N\int d\theta \,\alpha = 2\pi N \alpha.
\eea
This formula shows in particular that the magnetic flux is not
quantised. Mathematically this means that the gauge field $(A_1,A_2)$
of solutions of BVP does not extend  to the compactification
{\bf R}$^2 \cup \{\infty\}$; for if it did, $\Phi/(2\pi)$ would be
the first Chern number of a $U(1)$ bundle over a compact manifold,
which is necessarily an integer. The magnetic flux is nonetheless an
interesting quantity to consider because it  is  conserved
if one rules out infinite energy configurations. This follows from Faraday's
law of induction
\bea
{d \Phi \over d t} = \oint_C \bE\cd d \bl,
\eea
where the contour $C$  is the circle at
infinity.  The integral on the right hand side is only non-zero
if the electric field $\bE$ falls off  for large  $r$    no  faster than
$1/r$, which is precisely the condition for the  electric field to
have infinite  energy.

We have numerically solved BVP  with $N=2$ for various values  of
$\Phi$ in the allowed range $(-2\pi, 0)$.
 Since the qualitative features of  the
functions $f$ and  $a$  are clear from the proof of the proposition
we only show plots of  the  energy density and the magnetic field of
the solutions. The energy density is maximal
on a ring whose radius  is another useful measure of the soliton  size.
As the
magnitude of the  magnetic flux increases
this radius increases and reaches a finite limit for
 $|\Phi| \rightarrow 2\pi$.
In Figs. 1 and 2 the energy density and
the  magnetic field for a soliton whose (modulus of the) magnetic flux
is close to that limit are drawn with a solid line. In the limit
 $|\Phi| \rightarrow 0$ the soliton's size becomes arbitrarily small.

\section{Discussion and outlook}
We have computed rotationally symmetric solutions
of the  Bogomol'nyi equations in a
gauged version of the $O(3)$ sigma model  of degree $N$  with  $|N|\neq 1$.
There is no finite energy solution of degree 1,  but
there are  probably many more solutions of degree $N>1$ than considered here.
As in other field theories of Bogomol'nyi type
this can presumably be shown  using  an index theorem and a vanishing
theorem for an appropriate Dirac operator.
 Typically, there is a whole  manifold of
degree $N$ solutions of the Bogomol'nyi equation
 (called a moduli space), and the
dimension of this manifold is a linear function of $N$.

Solitons of Bogomol'nyi type which
display  all these properties and which are rather
similar to the solitons discussed here  are the
 $Q$-lumps mentioned in the introduction.
$Q$-lumps of degree 1 necessarily have infinite energy, but
there exists a  $(4N-2)$-dimensional  family of $Q$-lumps of degree
$N>1$. These include configurations which are made up of $N$
well-separated single $Q$-lumps. Similarly  there should
be  solutions of degree $N>1$  in the present  model
 whose  energy
density is peaked  at $N$  points in the plane.  These
could then be interpreted as superpositions of $N$  solitons
of degree 1; in this sense, solitons of degree $1$
can exist  as part of a multisoliton configuration.

Like $Q$-lumps the solitons discussed here can have an arbitrary
size, with the role
 of the size parameter   being played by the magnetic flux.
However, whereas the energy of a  $Q$-lump varies with the $Q$-lump's size,
the energy of the solitons discussed here is degenerate
with respect to changes  in the magnetic flux. Thus
the scaling degeneracy of the  pure  $O(3)$ sigma model persists
in  the present
model in a mutated form. There is an important difference, however.
In the $O(3)$ sigma model,  solitons exhibit a
 ``rolling instability", in the  sense that under a small perturbation
they either shrink to a thin spike or expand without limit (there is a
subtlety here: in the so-called moduli space approximation such scale
changes require infinite energy for single solitons
 and most multisolitons; numerical
simulations, however,  suggest that
they do occur in the full field theory).
 In  the present  model, by contrast,
there is  a completely general argument
 - Faraday's law of induction -    according
to which the total flux of a  configuration can only change at the
cost of infinite energy.

The constancy of the total flux does not prevent individual solitons
in a multisoliton configuration  from shrinking into thin spikes. It
would be interesting to see if this happens as a consequence of
soliton interactions. If the above conjecture about moduli spaces for
solitons of degree $N>1$ is correct this question could be investigated
using the moduli space approximation to soliton dynamics.

Finally  we  return to the question of
stabilising solitons in the $O(4)$ sigma model in (3+1) dimensions
with a  Maxwell term. This would lead to the  ``Skyrme-Maxwell
 model without a Skyrme term" mentioned in the introduction.
Of course it is well-known that
topologically non-trivial  gauge fields can stabilise solitons in
2 or 3 spatial dimensions,
like in the  case of abelian Higgs vortices or
t'Hooft-Polyakov  magnetic monopoles.
However, the
the analysis of the gauged $O(3)$ sigma model  here indicates that
such a stabilisation is also possible if there is only a dynamical
reason why the gauge  field cannot vanish during time evolution.
Thus one could  try to  stabilise a Skyrmion in
three spatial dimensions  via electric charge. The electric charge is
conserved  and, if it is non-zero,  produces an  electric
field  which prevents the Skyrmion from collapsing to zero size.
It seems quite possible that there are such electrically
charged Skyrmion solutions
in (3+1)-dimensional ``Skyrme-Maxwell theory without at Skyrme term".
The basic idea is to balance the tendency of the scalar field to
collapse by  the electrostatic repulsion of like charges.

\vspace{1cm}

\noindent {\bf Acknowledgements}

\noindent I   thank Bernard Piette   for  useful discussions
and acknowledge an SERC research assitantship.

\pagebreak
\parindent 0pt
\centerline{\bf Figure Captions}
\vspace{1cm}
{\bf Fig. 1}. The energy density   as a function of $r$ for solitons of
degree $N=2$  with  $\Phi/(2\pi) = -0.96$ (solid line),
 $\Phi/(2\pi) =-0.43$
(dashed line) $\Phi/(2\pi) = -0.18$ (dashed-dotted line). The function
$e$ plotted here is the integrand of (\ref{gauge}) divided by $4\pi$.

\vspace{1cm}
{\bf Fig. 2}. The   magnitude of the magnetic field
 $F_{12}$ as a  function of $r$
for solitons of degree $N=2$
 with  $\Phi/(2\pi) = -0.96$ (solid line), $\Phi/(2\pi) =-0.43$
(dashed line) $\Phi/(2\pi) = -0.18$ (dashed-dotted line).

\vspace{1cm}

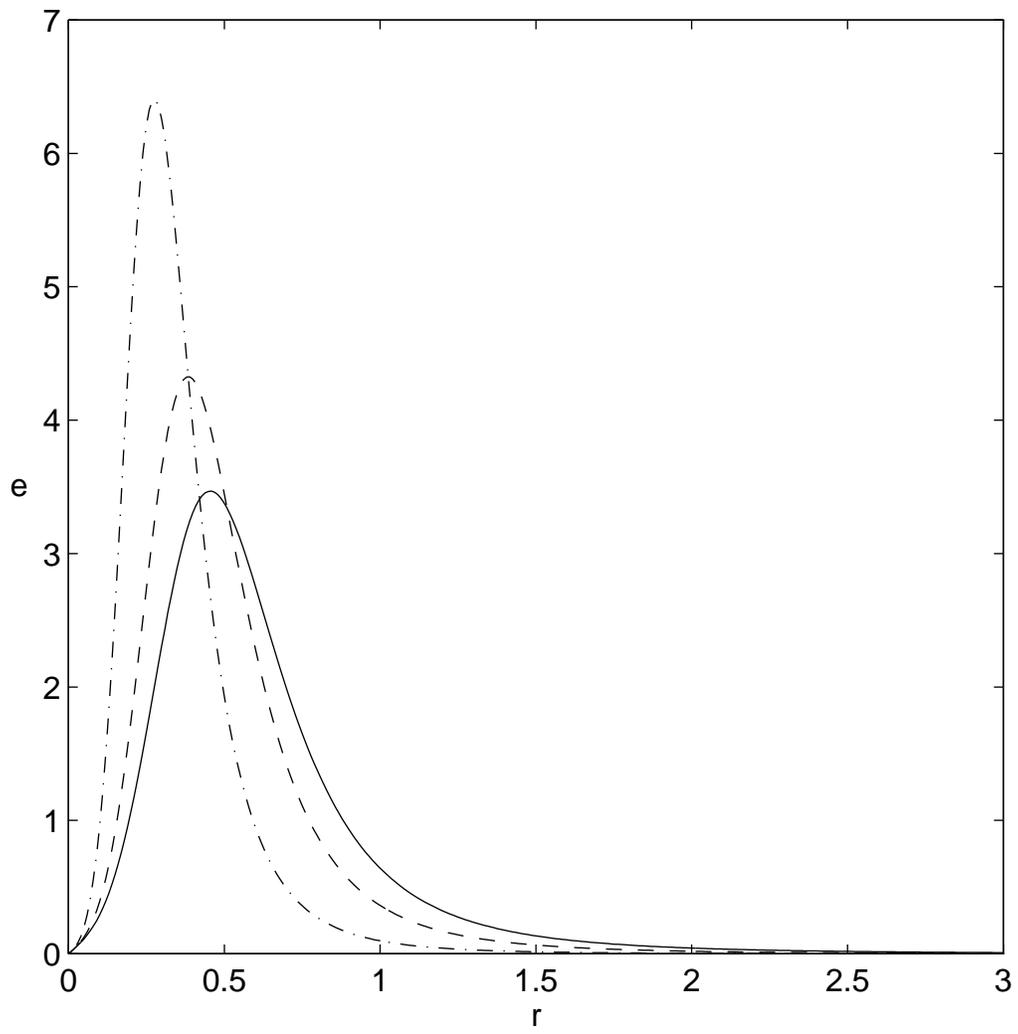
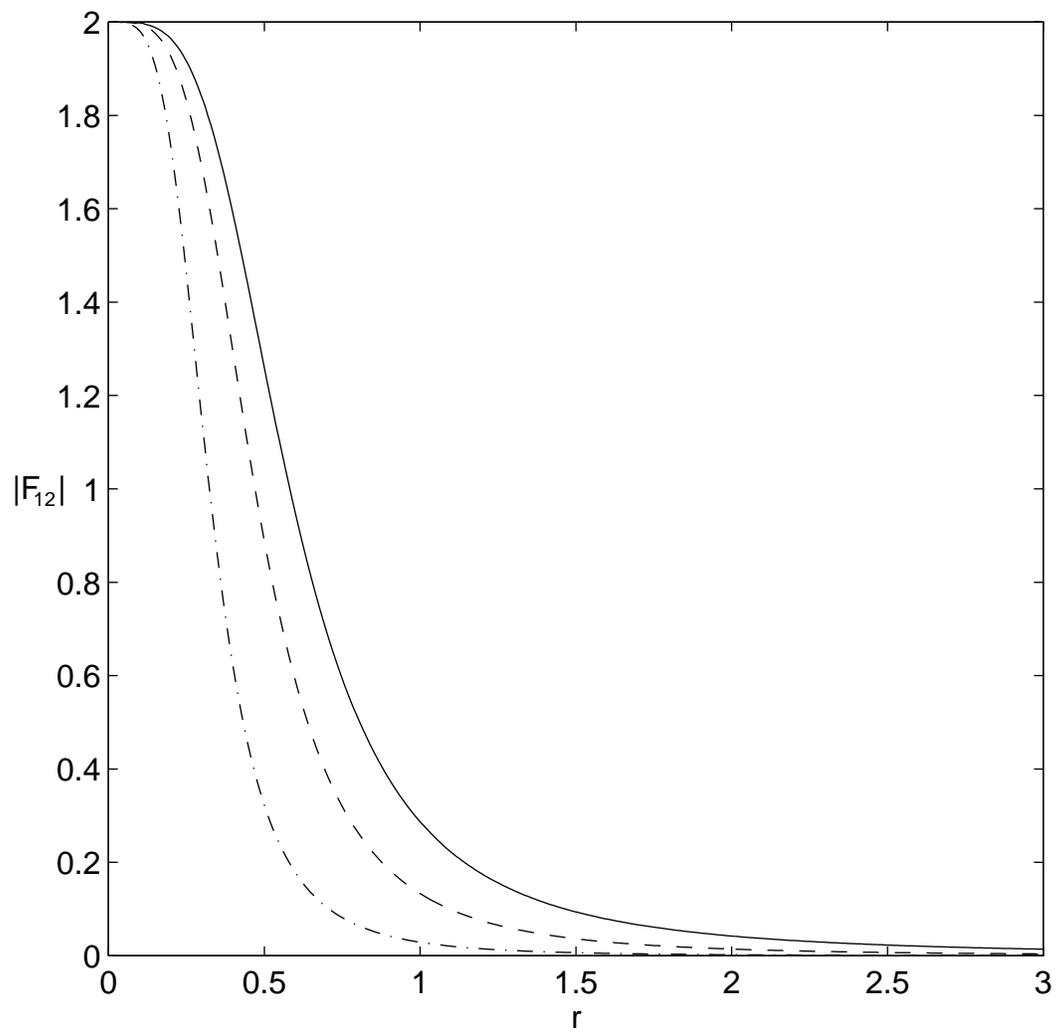
\end{document}